\documentclass[prd,twocolumn,showpacs,amsmath,amssymb]{revtex4-2}
\usepackage{subfigure}
\usepackage{graphicx}
\usepackage[colorlinks,linkcolor=blue,anchorcolor=blue,citecolor=blue]{hyperref}

\DeclareGraphicsExtensions{.pdf,.jpeg,.png}

\usepackage{lineno}

\newcommand{\ee}{e^+e^-}
\newcommand{\pip}{\pi^+}

\newcommand{\pim}{\pi^-}

\renewcommand*{\ee}{\ensuremath{e^+e^-}}

\newcommand*{\Dstar}{\ensuremath{D^{\ast 0}}}
\newcommand*{\Dstarpm}{\ensuremath{D^{\ast +}}}

\newcommand*{\DstarDee}{\ensuremath{D^{\ast 0}\to D^{0}e^{+}e^{-}}}
\newcommand*{\DsstarDee}{\ensuremath{D^{\ast +}_{s}\to D_{s}^{+}e^{+}e^{-}}}
\newcommand*{\DstarDg}{\ensuremath{D^{\ast 0}\to D^{0}\gamma}}
\newcommand*{\DsstarDg}{\ensuremath{D^{\ast +}_{s}\to D_{s}^{+}\gamma}}

\newcommand*{\Dstpair}{\ensuremath{D^{\ast 0}\bar{D}}{}\ensuremath{^{\ast 0}}}

\newcommand*{\eeDstpair}{\ee \ensuremath{\to} \Dstpair}

\newcommand*{\mBC}{\ensuremath{M_{\mathrm{BC}}}}
\newcommand*{\dE}{\ensuremath{\Delta E}}

\begin{document}
\normalsize
\parskip=5pt plus 1pt minus 1pt



\title{Observation of the electromagnetic Dalitz decay \DstarDee}

\author{
\begin{small}
\begin{center}
M.~Ablikim$^{1}$, M.~N.~Achasov$^{10,b}$, P.~Adlarson$^{66}$, S. ~Ahmed$^{14}$, M.~Albrecht$^{4}$, R.~Aliberti$^{27}$, A.~Amoroso$^{65A,65C}$, M.~R.~An$^{31}$, Q.~An$^{62,48}$, X.~H.~Bai$^{56}$
% ...
\\
\vspace{0.2cm}
(BESIII Collaboration)\\
\vspace{0.2cm} {\it
$^{1}$ Institute of High Energy Physics, Beijing 100049, People's Republic of China\\
% ...
$^{4}$ Bochum Ruhr-University, D-44780 Bochum, Germany\\
% ...
$^{10}$ G.I. Budker Institute of Nuclear Physics SB RAS (BINP), Novosibirsk 630090, Russia\\
% ...
$^{14}$ Helmholtz Institute Mainz, Staudinger Weg 18, D-55099 Mainz, Germany\\
% ...
$^{27}$ Johannes Gutenberg University of Mainz, Johann-Joachim-Becher-Weg 45, D-55099 Mainz, Germany\\
% ...
$^{31}$ Liaoning Normal University, Dalian 116029, People's Republic of China\\
% ...
$^{48}$ State Key Laboratory of Particle Detection and Electronics, Beijing 100049, Hefei 230026, People's Republic of China\\
% ...
$^{56}$ University of Jinan, Jinan 250022, People's Republic of China\\
% ...
$^{62}$ University of Science and Technology of China, Hefei 230026, People's Republic of China\\
% ...
$^{65}$ University of Turin and INFN, (A)University of Turin, I-10125, Turin, Italy; (B)University of Eastern Piedmont, I-15121, Alessandria, Italy; (C)INFN, I-10125, Turin, Italy\\
$^{66}$ Uppsala University, Box 516, SE-75120 Uppsala, Sweden\\
% ...
$^{b}$ Also at the Novosibirsk State University, Novosibirsk, 630090, Russia\\
% ...
}\end{center}
\vspace{0.4cm}
\end{small}
}
\date{\today}

\newpage

\begin{abstract}

 Based on 3.19 fb$^{-1}$ of $e^+e^-$ collision data accumulated at the center-of-mass energy  4.178~GeV 
 with the BESIII detector operating at the BEPCII collider, 
 the electromagnetic Dalitz decay \DstarDee{} is observed for the first time with a statistical significance of $13.2\sigma$. The ratio of the branching fraction of \DstarDee{} to that of \DstarDg{} is measured to be 
 $(11.08\pm0.76\pm0.49)\times 10^{-3}$. By using the world average value of the branching fraction of \DstarDg{}, 
  the branching fraction of \DstarDee{} is determined to be $(3.91\pm0.27\pm0.17\pm0.10)\times 10^{-3}$,
  where the first uncertainty is statistical, the second is systematic and the third is from external input of the branching fraction for \DstarDg{}.  
  
\end{abstract}

\date{\today}

\maketitle

\section{Introduction}

The conversion decays of a vector resonance ($V$) into a pseudoscalar meson ($P$) 
and a lepton pair, $V \to P\gamma^\ast \to P\ell\bar{\ell}$, provide a stringent test for theoretical 
models of the structure of hadrons and the interaction mechanism between photons and hadrons~\cite{Landsberg:1986fd}. In these processes, the squared dilepton
invariant mass corresponds to the virtual photon 4-momentum
transfer squared, $q^2$. The $q^2$ distribution depends on the underlying dynamical electromagnetic 
structure of the transition $V \to P\gamma^\ast$. Furthermore, through the  vector-meson dominance (VMD) model~\cite{Landsberg:1986fd}, the virtual photon effectively couples to vector particles, such as the $\rho$, $\omega$, $\phi$ and even the heavier $J/\psi$ particle. However, the off-shell effects cannot be calculated reliably~\cite{Wu:2019adv}. Experimental measurement of the $q^2$ distribution is crucial to validate these calculations.

\begin{figure}[h]
\begin{center}
\subfigure[]{\includegraphics[width=0.33\textwidth]{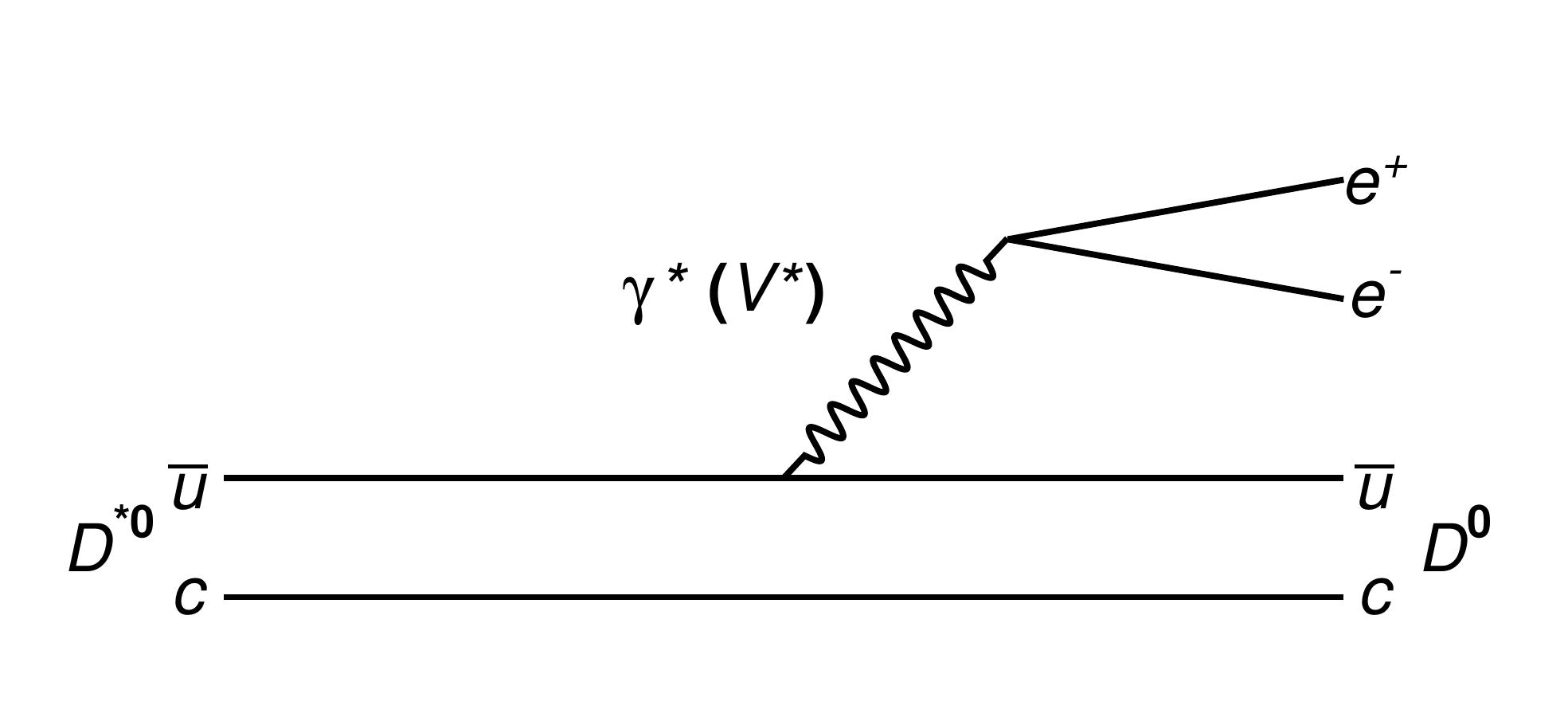}}
\subfigure[]{\includegraphics[width=0.33\textwidth]{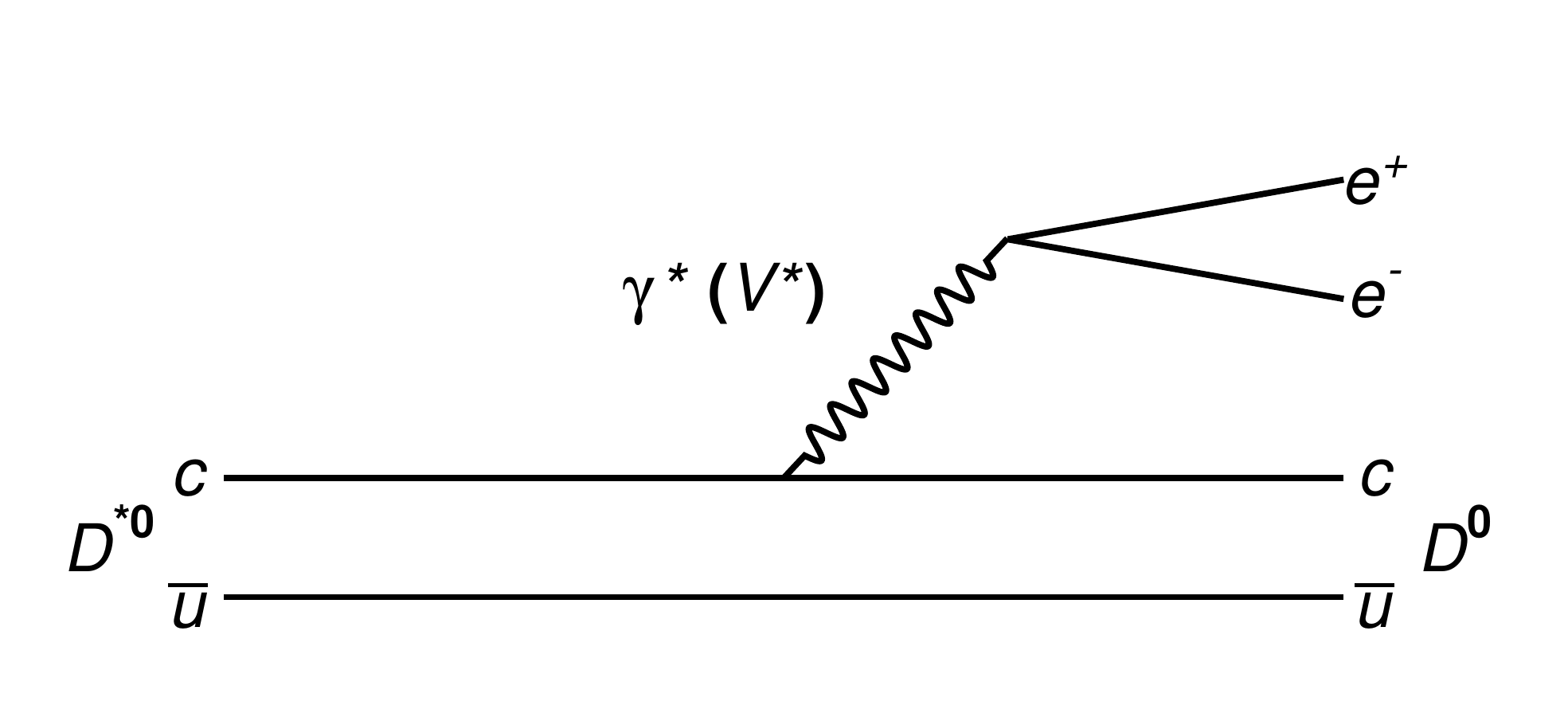}}
\end{center}
\caption{Diagrams of the decay \DstarDee. The symbol $V^*$ indicates the virtual $\rho$, $\omega$, $\phi$ or $J/\psi$ meson. } \label{fig:feynman}
\end{figure}

Several light meson decays with virtual photon-conversion to an $e^+e^-$ pair, 
e.g., $\omega\to \pi^0 e^+e^-$~\cite{Achasov:2008zz},
$\phi\to\pi^0 e^+e^-$~\cite{Anastasi:2016qga}, and $\phi\to \eta e^+e^-$~\cite{Babusci:2014ldz}, have been observed. 
The charmonium electromagnetic (EM) Dalitz decay, $\psi(3686)\to e^+e^-\eta^{\prime}$~\cite{Ablikim:2018xxs}, 
$\psi(3686)\to e^+e^-\chi_{cJ}$~\cite{Ablikim:2017kia}, and $J/\psi \to e^+e^-\eta$~\cite{BESIII:2018qzg}, 
have also been reported. The charm meson decay \DsstarDee{} was observed by CLEO~\cite{CroninHennessy:2011xp},  
and the ratio of the branching fractions $\frac{\mathcal{B}(\DsstarDee) } {\mathcal{B}(\DsstarDg)}$ is measured to be ($0.72\pm0.18$)\%, 
which agrees with theoretical calculation based on the VMD model. 
However, there is no experimental result for the corresponding EM Dalitz decays of \Dstar{} and \Dstarpm{}. 
According to the VMD model, the \DsstarDee{} decay mainly occurs via the coupling of the virtual photon to the vector $\phi$ meson, 
while the \DstarDee{} decay proceeds through the coupling to the $\rho$ or $\omega$ meson. 
Hence, the study of the decay \DstarDee{}, with the diagrams depicted in \figurename~\ref{fig:feynman},
provides information on the form factor for the couplings $\gamma^{*}\to\rho$ and $\gamma^{*}\to\omega$. 
The ratio between branching fractions of \DstarDee{} and \DstarDg{} is defined as 
\begin{eqnarray} \label{eq:Ree}
  R_{ee}=\frac{{\mathcal B}(\DstarDee)}{{\mathcal B}(\DstarDg)}.
\end{eqnarray}
Calculations using the VMD model give $R_{ee}=0.67\%$ along with the following differential decay rate~\cite{Landsberg:1986fd}: 
\begin{eqnarray} \label{Eq:2}
	\frac{dR_{ee}} {dq^{2}}&= &\frac{\alpha}{3\pi q^{2}} \left \lvert \frac{f(q^{2})}{f(0)} \right \rvert^{2} \left[1-\frac{4m_{e}^{2}}{q^{2}}\right]^{\frac{1}{2}}\left[1+\frac{2m_{e}^{2}}{q^{2}}\right] \nonumber \\ 
& &	\times \left[\left(1+\frac{q^{2}}{A}\right)-\frac{4m_{D^{*0}}^{2}q^{2}}{A^{2}}\right]^{\frac{3}{2}}. 
\end{eqnarray}
Here, $\alpha$ is the fine structure constant, $A = m_{D^{*0}}^{2} - m_{D^{0}}^{2}$, 
$f(q^{2})$ is the transition form factor for \Dstar{}  to $D^{0}$, $m_e$ is the  mass of electron and $m_{D^{*0}}$ 
 ($m_{D^{0}}$) is the mass of $D^{*0}$ ($D^{0}$).  
The form-factor ratio $\frac{f(q^{2})}{f(0)}$ is equal to $(1-\frac{q^{2}}{m_{\rho}^{2}})^{-1}$, where $m_{\rho}$ is the $\rho$ resonance mass.

In this paper, the EM Dalitz decay \DstarDee{} is studied and $R_{ee}$ is measured using 3.19 fb$^{-1}$ of $e^+e^-$ collision data 
collected with the BESIII detector at the center-of-mass energy $\sqrt{s}=4.178$~GeV.  
To control background contributions, the candidates for both \DstarDee{} and \DstarDg{} are reconstructed with \eeDstpair{} events. The single tag method is used, so that only  $D^{*0}$ ($\bar{D}^{*0}$) is reconstructed and the $\bar{D}^{*0}$($D^{*0}$) on the recoiling side is not detected.
The inclusion of charge-conjugate states is implied throughout this paper.

\section{Description of the BEPCII and the BESIII detector}

The BESIII detector~\cite{Ablikim:2009aa} records symmetric $e^+e^-$ collisions 
provided by the BEPCII storage ring~\cite{Yu:IPAC2016-TUYA01}, which operates in the center-of-mass energy range from $2.0-4.96$~GeV.
BESIII has collected large data samples in this energy region~\cite{BESIII:2020nme}. The cylindrical core of the BESIII detector covers 93\% of the full solid angle and consists of a helium-based
 multilayer drift chamber~(MDC), a plastic scintillator and multi-gap resistive plate chamber time-of-flight 
system~(TOF), and a CsI(Tl) electromagnetic calorimeter~(EMC),
which are all enclosed in a superconducting solenoidal magnet
providing a 1.0~T magnetic field. The solenoid is supported by an
octagonal flux-return yoke with resistive plate counter muon
identification modules interleaved with steel. 
The charged-particle momentum resolution at $1~{\rm GeV}/c$ is
$0.5\%$, and the $dE/dx$ resolution is $6\%$ for electrons
from Bhabha scattering. The EMC measures photon energies with a
resolution of $2.5\%$ ($5\%$) at $1$~GeV in the barrel (end cap)
region. The time resolution in the TOF barrel region is 68~ps, while
that in the end cap region is 60~ps~\cite{etof}.

\section{Monte Carlo simulation}
Simulated data samples produced with a {\sc
geant4}-based~\cite{geant4} Monte Carlo (MC) package, which
includes the geometric description of the BESIII detector and the
detector response, are used to determine detection efficiencies
and to estimate backgrounds. The simulation models the beam
energy spread and initial state radiation (ISR) in the $e^+e^-$
annihilations with the generator {\sc kkmc}~\cite{ref:kkmc}. 
The inclusive MC sample includes the production of open charm
processes, the ISR production of vector charmonium(-like) states,
and the continuum processes incorporated in {\sc
kkmc}~\cite{ref:kkmc}. The known decay modes are modeled with {\sc
evtgen}~\cite{ref:evtgen} using branching fractions taken from the
Particle Data Group \cite{Zyla:2020zbs}, and the remaining unknown charmonium decays
are modeled with {\sc lundcharm}~\cite{ref:lundcharm}. Final state radiation~(FSR)
from charged final state particles is incorporated using {\sc
photos}~\cite{photos}.

The MC events of the signal process \eeDstpair{}, \DstarDee{} are generated according to the VMD model. Following Ref.~\cite{Tan:2021clg}, the decay \DstarDee{} is described as
\begin{eqnarray} \label{Eq:3}
	\frac{d\Gamma}{dq^{2}d\cos{\theta_e}} & \propto &  \frac{|f(q^{2})|^{2}}{q^{2}}\left(1-\frac{4m^2_{e}}{q^{2}}\right)^{\frac{1}{2}} \nonumber \\
	&\times & [(m_{D^{*0}}^{2} - m_{D^0}^2 + q^{2})^2 - 4m_{D^{*0}}^{2}q^{2}]^{\frac{3}{2}} \nonumber \\
       &\times & \left[\left(1+\frac{4m_{e}^2}{q^2}\right)+\left(1-\frac{4m_{e}^2}{q^2}\right)\cos^2{\theta_e}\right],
\end{eqnarray}
where $\theta_e$ is the helicity angle of the electron pair system. 
The MC samples of the reference process \DstarDg{} are generated with the {\sc kkmc}.

\section{Event selection}
To reconstruct the signal and reference processes,  $D^{0}$ candidates are selected via
three decay modes $D^{0}\rightarrow K^{-}\pi^{+}$, $K^{-}\pi^{+}\pi^{0}$ and $K^{-}\pi^{+}\pi^{-}\pi^{+}$.
For each charged track candidate, the polar angle $\theta$ in the MDC is required to be in the range $|\!\cos\theta|<0.93$, and the distance of closest approach to the interaction point is required to be
less than 10~cm along the beam direction and less than 1~cm in the
plane perpendicular to the beam.  The $dE/dx$ recorded by the MDC and the time-of-flight information measured by the TOF are combined to calculate particle identification (PID) probability for the pion 
($\mathcal{P}_{\pi}$) and kaon ($\mathcal{P}_{K}$) hypotheses. Pion candidates are selected by requiring $\mathcal{P}_{\pi} > 0$ and $\mathcal{P}_{\pi} > \mathcal{P}_{K}$, and kaon candidates are required to satisfy $\mathcal{P}_K > 0$ and $\mathcal{P}_K> \mathcal{P}_{\pi}$.
Photon candidates are reconstructed with isolated clusters in the
EMC in the region $|\!\cos\theta| \le 0.80$ (barrel) or $0.86 \le |\!\cos\theta| \le 0.92$ (end cap). The deposited energy of the cluster is required to be larger than 25 (50)~MeV in the barrel (end cap) region, and the angle between the photon candidate and any charged track is larger than 10$^\circ$. 
All $\gamma\gamma$ combinations are considered as candidate $\pi^{0}$ mesons,
 and the reconstructed mass $M_{\gamma\gamma}$ is required to satisfy 
 $0.115<M_{\gamma\gamma}<0.150$~GeV/$c^{2}$. A kinematic fit is performed to constrain the $\gamma\gamma$ invariant mass to the
nominal $\pi^0$ mass taken from the PDG~\cite{Zyla:2020zbs}, 
and candidates with the fit quality $\chi^2<200$ are retained.
The $K^-\pi^+$, $K^-\pi^+\pi^0$, and $K^-\pi^+\pi^-\pi^+$ combinations are required to be within the mass windows 1.85$<M_{K^{-}\pi^{+}}<1.88$~GeV/$c^{2}$, 1.84$<M_{K^{-}\pi^{+}\pi^{0}}<1.89$~GeV/$c^{2}$ and 1.85$<M_{K^{-}\pi^{+}\pi^{-}\pi^{+}}<1.88$~GeV/$c^{2}$, respectively. For each decay mode, all possible combinations are kept for further analysis. 

\begin{figure*}[htp]
    \begin{center}
       \includegraphics[width=0.45\linewidth]{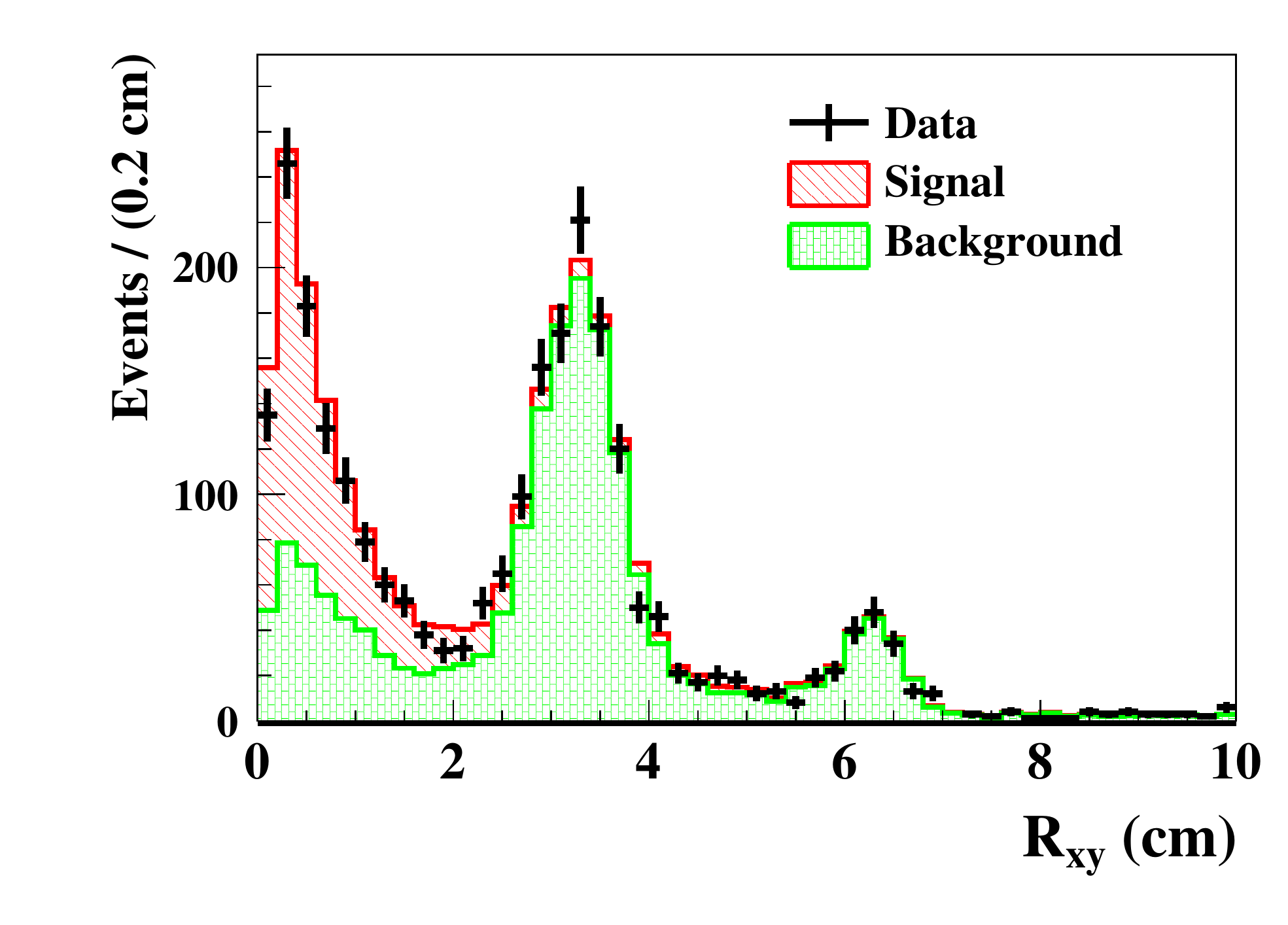}\label{fig:gamma_conversion_1D}
        \put(-110,140){(a)}
        \includegraphics[width=0.45\linewidth]{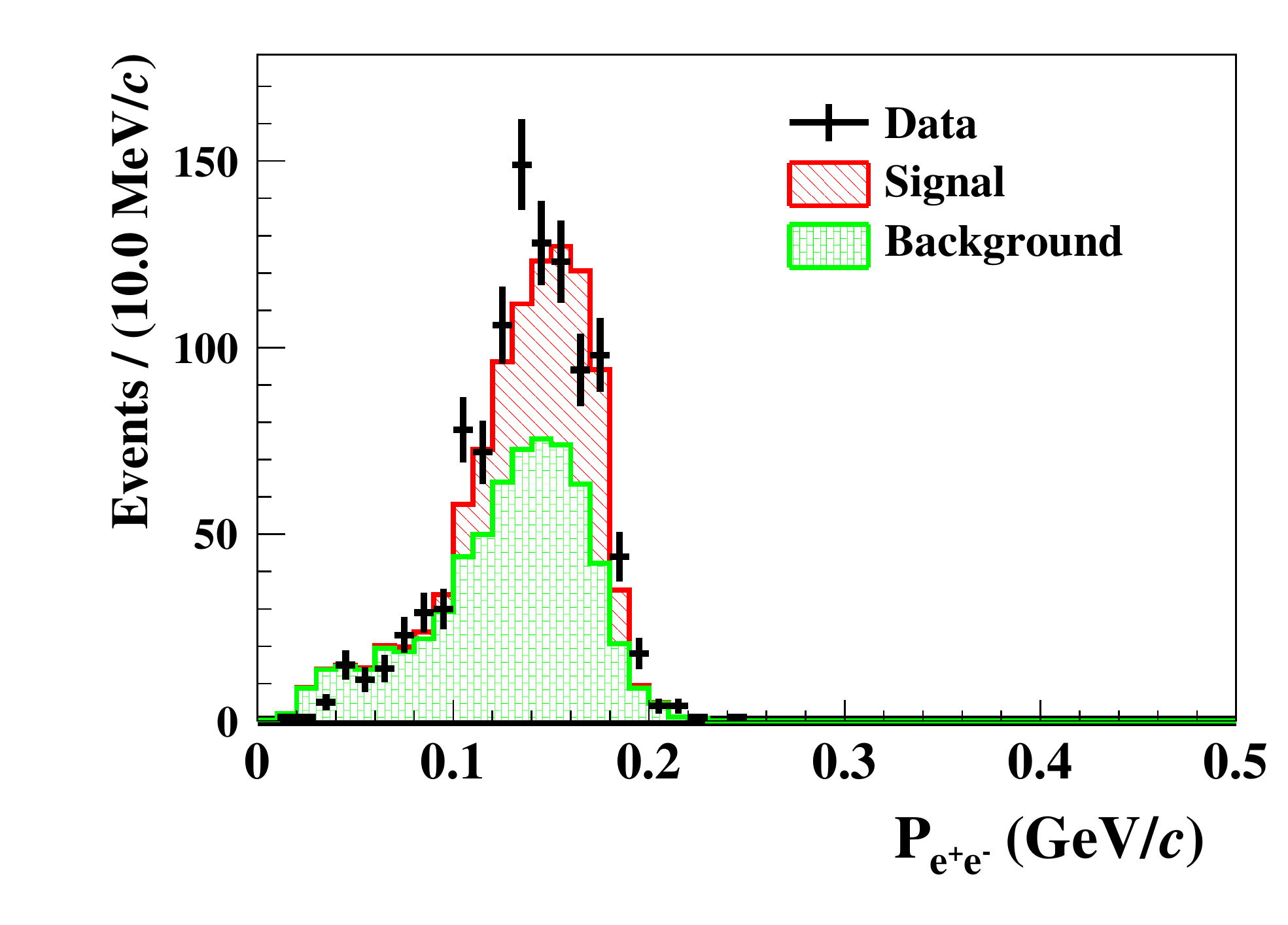}\label{fig:momentum_ee}
         \put(-110,140){(b)}

        \includegraphics[width=0.45\linewidth]{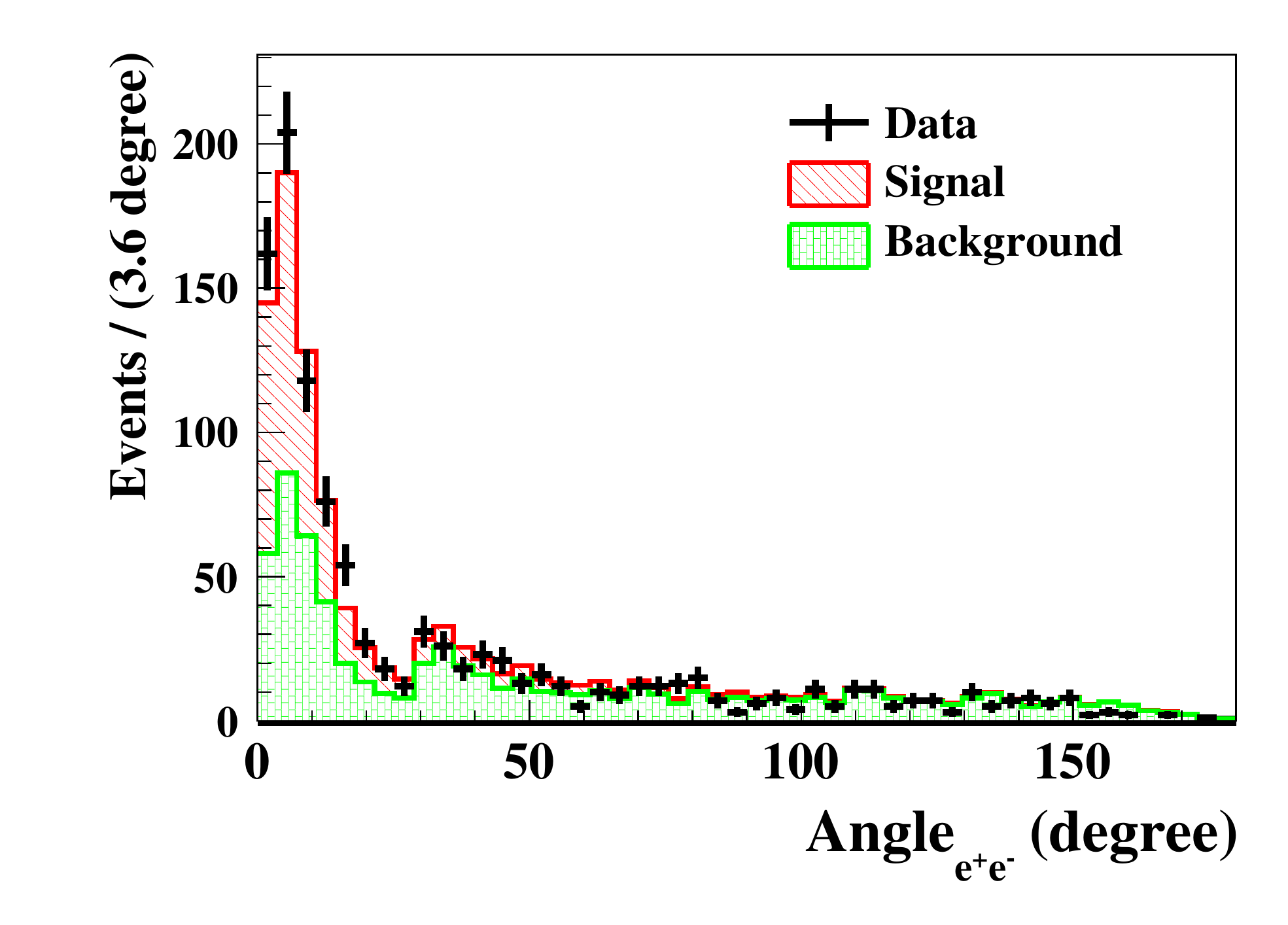}\label{fig:angle_ee}
          \put(-110,140){(c)}
         \includegraphics[width=0.45\linewidth]{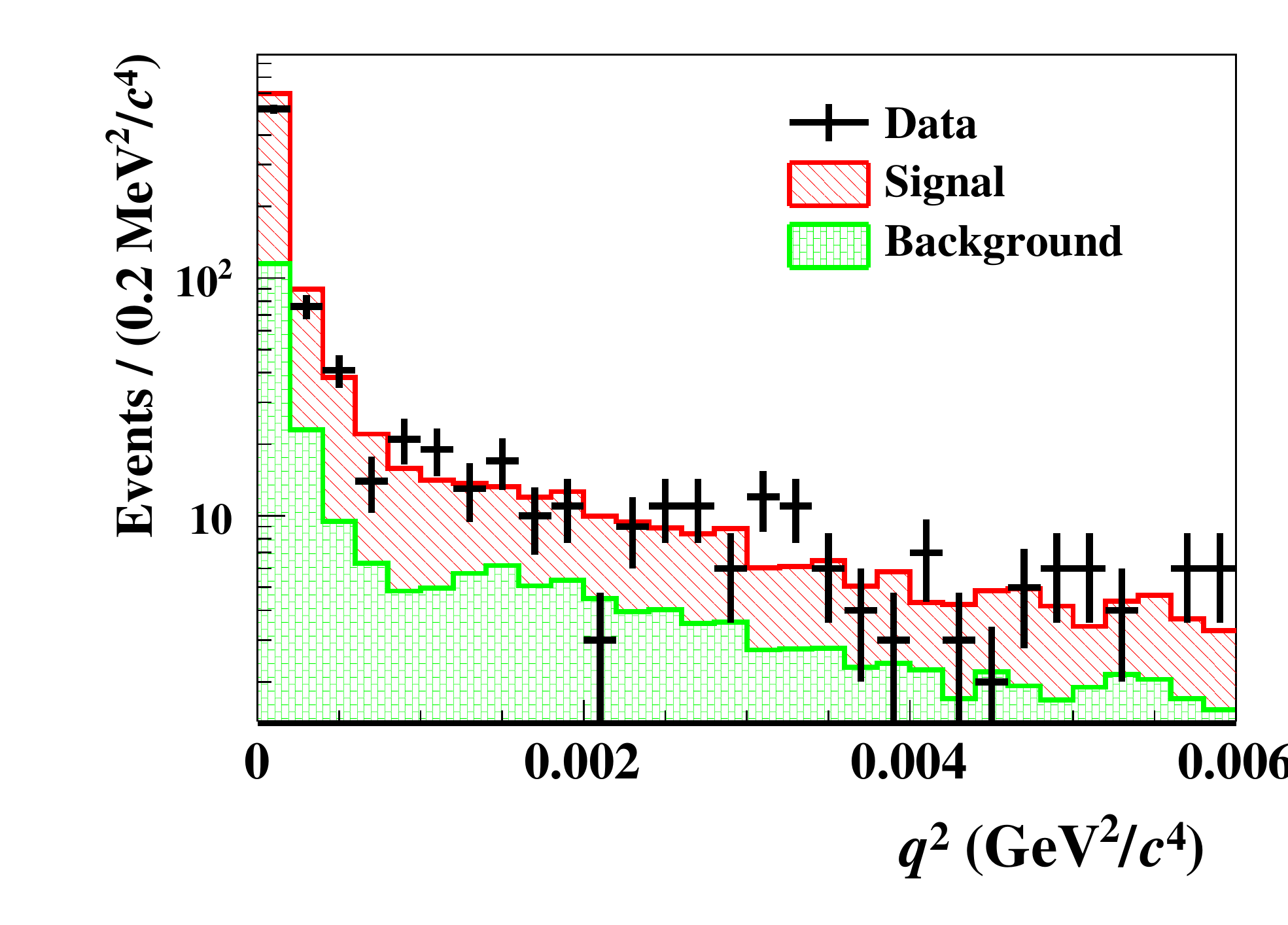}\label{fig:mee}
            \put(-110,140){(d)}
        \end{center}
    \caption{The distributions of (a)  $R_{xy}$,  
              (b) the momentum of the $e^{+}e^{-}$ pair,
                 (c) the angle between $e^{+}$ and $e^{-}$ in the laboratory frame and 
                 (d) the virtual photon 4-momentum transfer square ($q^2$) of the candidates for $D^{*0}\to D^0e^+e^-$. The  points with error bars are data.  The brick-filled green histograms indicate the scaled backgrounds derived from inclusive MC samples. The slash-filled red histograms label the normalized signal \DstarDee{} contributions, extracted from the signal MC samples. For all distributions, the three $D^{0}$ decay modes are combined and \mBC{} is required to be in the region (2.00, 2.02) GeV/$c^2$. The distributions of (b), (c) and (d) show events passing the $R_{xy} < 2.0$ cm requirement.
     } \label{fig:com_data_mc}
\end{figure*}

\begin{figure*}[htp]
    \begin{center}
       \includegraphics[width=0.32\textwidth,height=0.2\textheight]{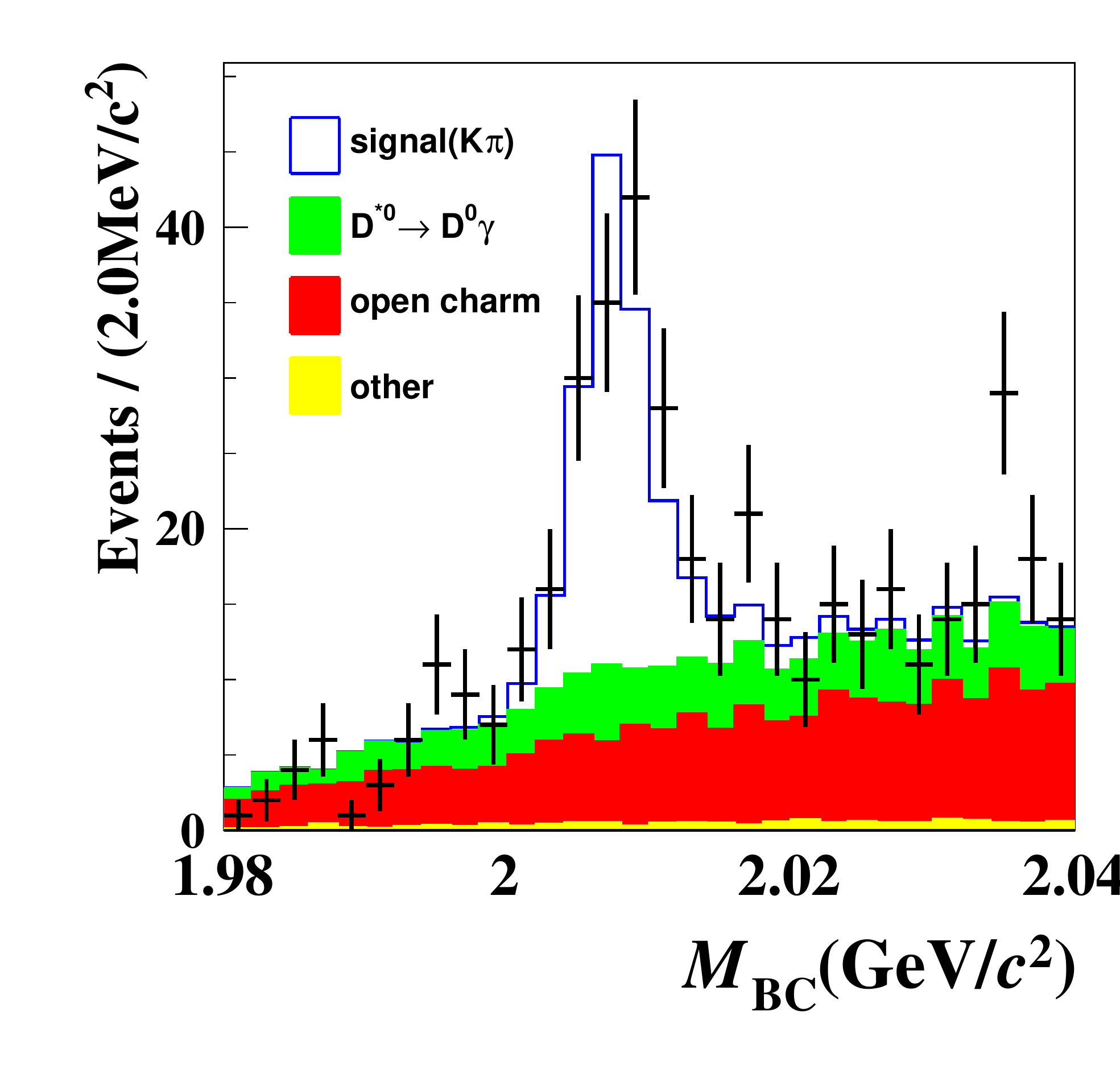}
  \put(-120,70){(a)}
       \includegraphics[width=0.32\textwidth,height=0.2\textheight]{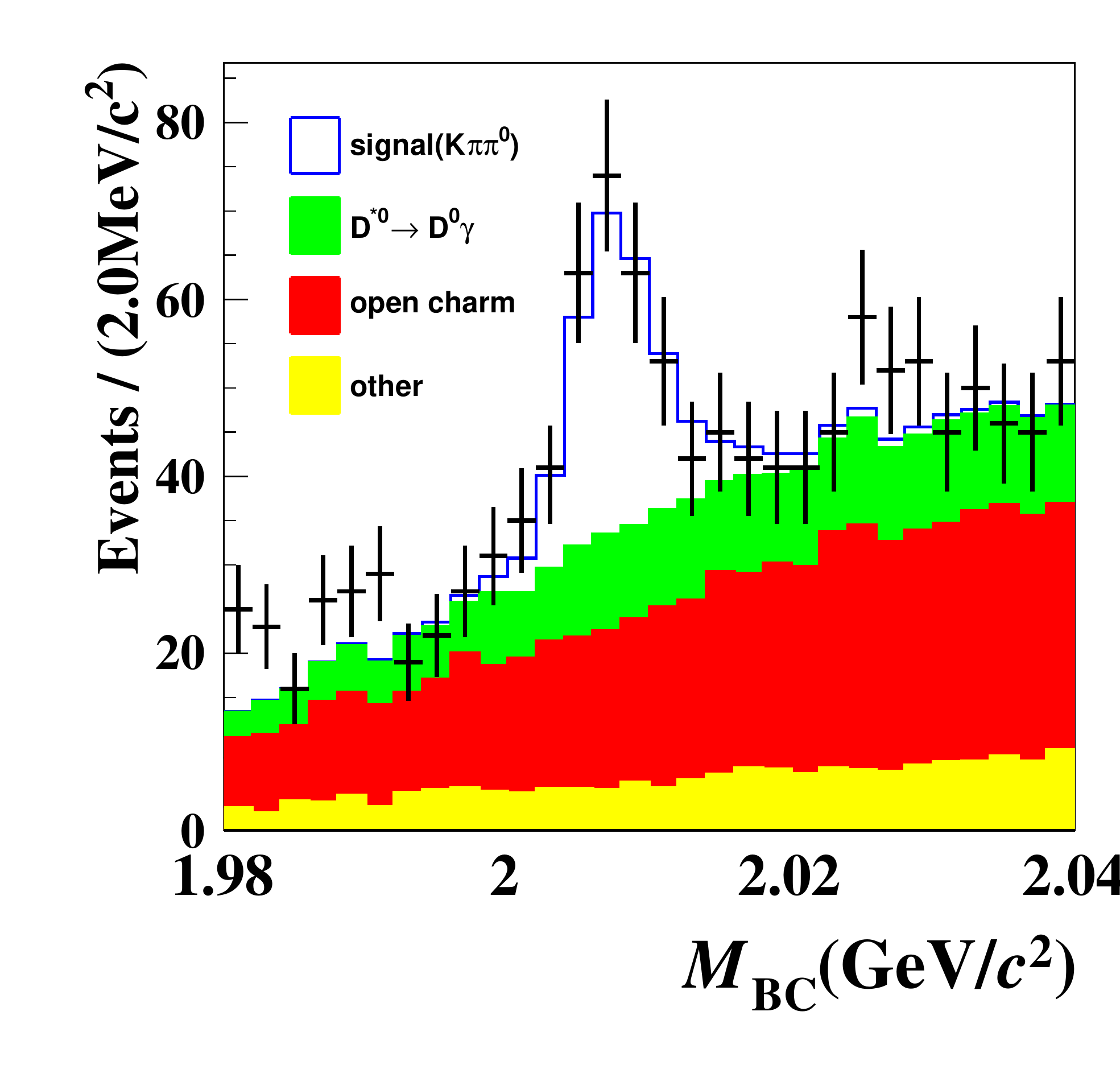}
  \put(-120,70){(b)}
       \includegraphics[width=0.32\textwidth,height=0.2\textheight]{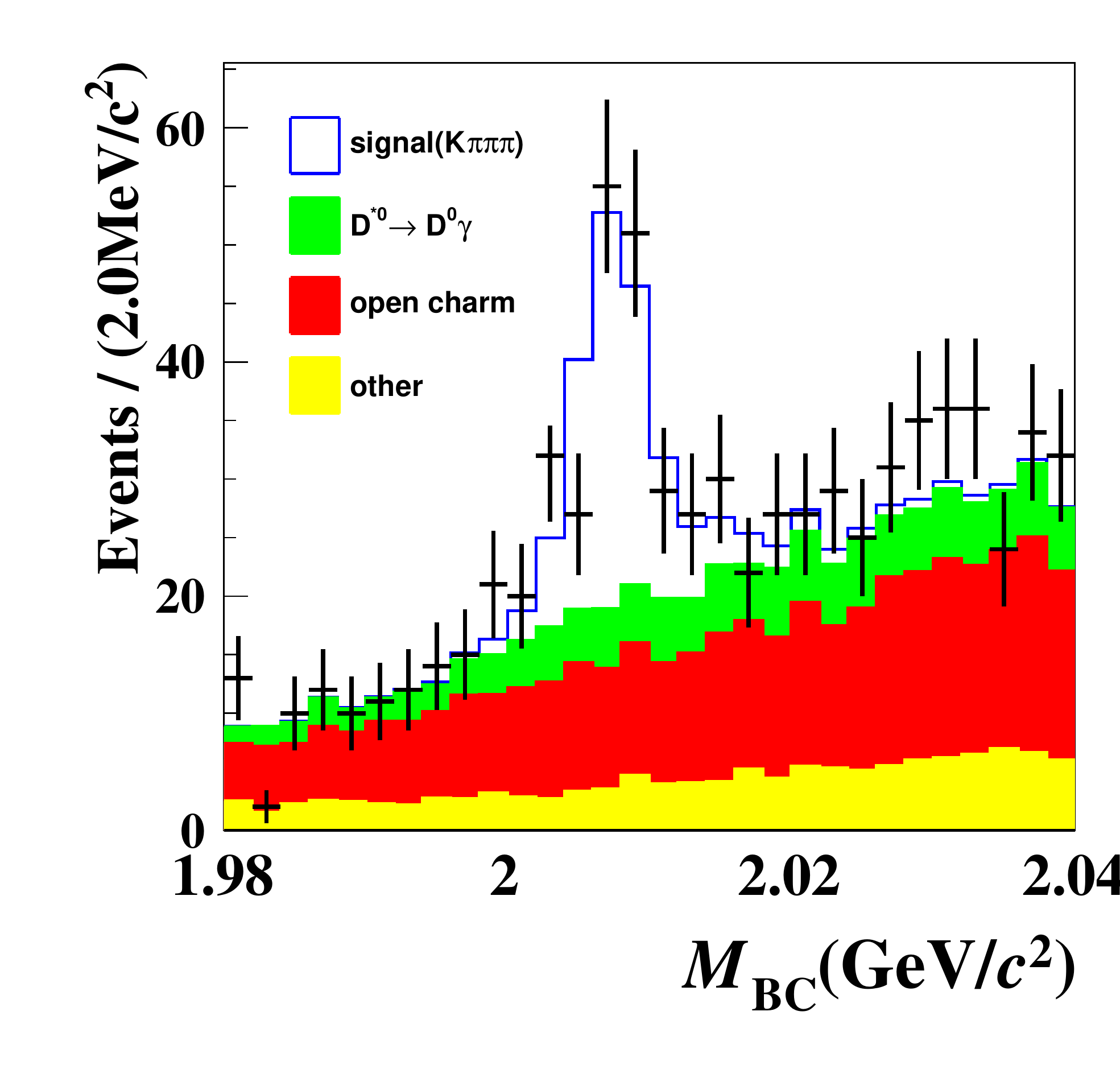}
  \put(-120,70){(c)}
        \end{center}
    \caption{The comparison between data and MC components for $D^{*0} \rightarrow D^{0} e^{+} e^{-}$ process, where plot (a) is for $D^{0}\rightarrow K^{-}\pi^{+}$ mode, plot (b) for $D^{0}\rightarrow K^{-}\pi^{+}\pi^{0}$ mode and plot (c) for $D^{0}\rightarrow K^{-}\pi^{+}\pi^{-}\pi^{+}$ mode. Black dots with error bars are data, the blue lines are signal processes. The shaded histograms are the background processes for $D^{*0} \rightarrow D^{0} \gamma$ (green), open charm (red) and other (yellow) processes, respectively.} 
 \label{fig:component}
\end{figure*}

For the \DstarDee{} decay, the electron and positron candidates are identified among the remaining charged tracks with the $dE/dx$ information measured by the MDC. 
The probability criteria of $\mathcal{P}_e>0$, $\mathcal{P}_e> \mathcal{P}_{\pi}$ and 
$\mathcal{P}_{e} > \mathcal{P}_{K}$ are applied.~The energy difference \dE{} is defined as  $\dE = E_{\Dstar} - E_{\rm beam}$,
where $E_{\Dstar}$ is the reconstructed energy of \Dstar{} candidates in the rest frame of the $\ee$ initial beams and $E_{\rm beam}$ is the beam energy; 
$|\dE|<0.03$~GeV is required to reduce background contributions.  
If there are multiple candidates (13\% of the selected events) in an event, only the one with the minimum $|\dE{}|$ is accepted.  
Similarly, for the \DstarDg{} candidates, if there are multiple combinations (32\% of the selected events) in an event, only the candidate with the minimum $|\Delta E|$ is kept and $|\dE|<0.03$~GeV is required to reduce backgrounds.
To separate $D^{*0}$ candidates produced from $e^+e^-\to D^{*0}\bar D^{*0}$ from backgrounds, the beam-constrained mass \mBC{} is defined as $\mBC^{2} = E_{\rm beam}^{2}-p_{\Dstar}^{2}$,
where $p_{\Dstar}$ is the measured total momentum of \Dstar{} candidates in the rest frame of initial $\ee$ beams. 
In addition, a veto for dielectrons from photon conversion is applied to suppress background from \DstarDg{},
where the dielectron comes from the transition photon interacting with the materials in the beam pipe and the  MDC inner wall. 
The variable of  $R_{xy}$ which represents the distance between beam interaction point and vertex of photon-conversion in $xy$ plane~\cite{Xu:2012xq} is calculated. Figure~\ref{fig:com_data_mc}(a) shows the $R_{xy}$ distribution of the  \DstarDee{} candidates, where two clear peaks corresponding to the beam pipe position (3 cm) and the MDC inner wall (6 cm) are observed. 
An additional requirement $R_{xy}<2.0$ cm  removes the photon-conversion events. Figures~\ref{fig:com_data_mc}(b), 2(c) and 2(d) show the distributions of the momentum of the $e^{+}e^{-}$ pair, the opening angle between $e^{+}$ and $e^-$, and the $q^2$ of the $e^{+}e^{-}$ candidates for the $D^{*0}\to D^0e^+e^-$, respectively. 
Good agreement between data and  MC simulation is seen in \figurename~\ref{fig:component}.

\section{Determination of the branching fraction}

In the data sample, the observed number of signal events is expressed as
\begin{align} \label{fig:FS}
N_{\rm sig} =2\cdot N_{\Dstpair{}}\cdot \mathcal{B}(\DstarDee)\cdot \mathcal{B}_{\mathrm{int}} \cdot \varepsilon_{\rm sig}, 
\end{align}
and the observed number of the reference process as
\begin{align} \label{fig:FF}
	N_{\rm ref}= 2\cdot N_{\Dstpair{}}\cdot \mathcal{B}(\DstarDg)\cdot \mathcal{B}_{\mathrm{int}}\cdot \varepsilon_{\rm ref}.
\end{align}
Here, $N_{D^{*0}\bar D^{*0}}$ is the total number of $D^{*0}\bar D^{*0}$ pairs in data, $\varepsilon_{\rm sig}$ and $\varepsilon_{\rm ref}$ denote the detection efficiencies of the signal and reference processes, respectively, and $\mathcal{B}_{\mathrm{int}}$ stands for the branching fractions for the three $D^{0}$ decay modes and secondary decay of the $\pi^0$ meson. For the signal process, the efficiency $\varepsilon_{\rm sig}$ has been corrected to account for differences of the photon conversion, tracking and PID efficiencies between data and MC simulation (see discussion of systematic uncertainties below). Thus, the ratio $R_{ee}$ for each decay mode is given as
\begin{eqnarray} \label{eq:ratio}
 R_{ee}  = \frac{{\mathcal B}(D^{*0}\to D^0e^+e^-)}{{\mathcal B}(D^{*0}\to D^0\gamma)}
     = \frac{N_{\rm sig}\cdot \epsilon_{\rm ref}}{N_{\rm ref}\cdot \epsilon_{\rm sig}}.
\end{eqnarray}

\begin{table*}[htp] 
\begin{center}

\caption{Yields and efficiencies of the three $D^{0}$ tag decay modes and the obtained branching fractions. For the obtained branching fractions, the first uncertainties are statistical and the second systematic, while the uncertainties are statistical only for the other numbers.  The third systematic uncertainty is quoted from the uncertainty of $D^{*0}\rightarrow D^{0}\gamma$ in PDG.} \label{tab:results}
   \begin{tabular}{l|rrr}
	\hline\hline
	 & $D^{0}\rightarrow K^{-}\pi^{+}$ & $D^{0}\rightarrow K^{-}\pi^{+}\pi^{0}$ & $D^{0}\rightarrow K^{-}\pi^{+}\pi^{-}\pi^{+}$\\
	\hline
     $\epsilon_{\rm ref}$ (\%) &  35.82$\pm$0.28& 14.81$\pm$0.20  &19.11$\pm$0.15         \\
      	$N_{\rm ref}$ & 66484$\pm$223\; &97471$\pm$327\; &74196$\pm$249\;  \\
      $\epsilon_{\rm sig}$ (\%) &5.61$\pm$0.04& 2.58$\pm$0.03  &3.09$\pm$0.03  \\
	$N_{\rm sig}$ & 111.3$\pm$7.6 \,  &181.5$\pm$12.4 & 128.1$\pm$8.7 \,  \\
	\hline
	$R_{ee}$ & \multicolumn{3}{c}{$(11.08\pm0.76\pm0.49)\times 10^{-3}$}\, \\
	$\mathcal{B}(\DstarDee)$ & \multicolumn{3}{c}{$(3.91\pm0.27\pm0.17\pm0.10)\times 10^{-3}$} \\
	\hline
	\hline
   \end{tabular}
\end{center}
 \end{table*}

\begin{figure*}[htp]
    \begin{center}
       \includegraphics[width=0.32\textwidth,height=0.2\textheight]{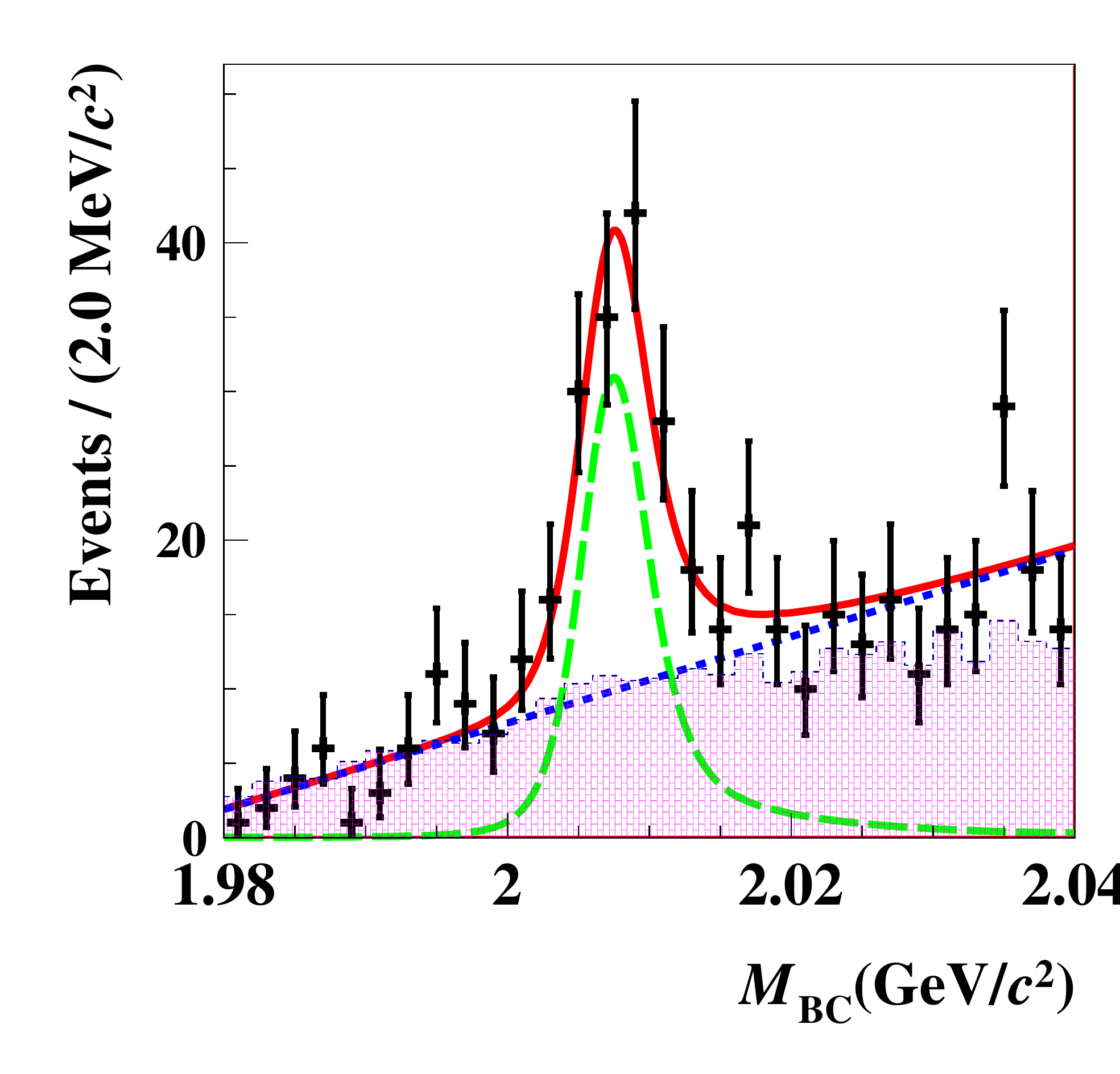}\label{fig:mBC_0}
  \put(-110,100){(a)}
       \includegraphics[width=0.32\textwidth,height=0.2\textheight]{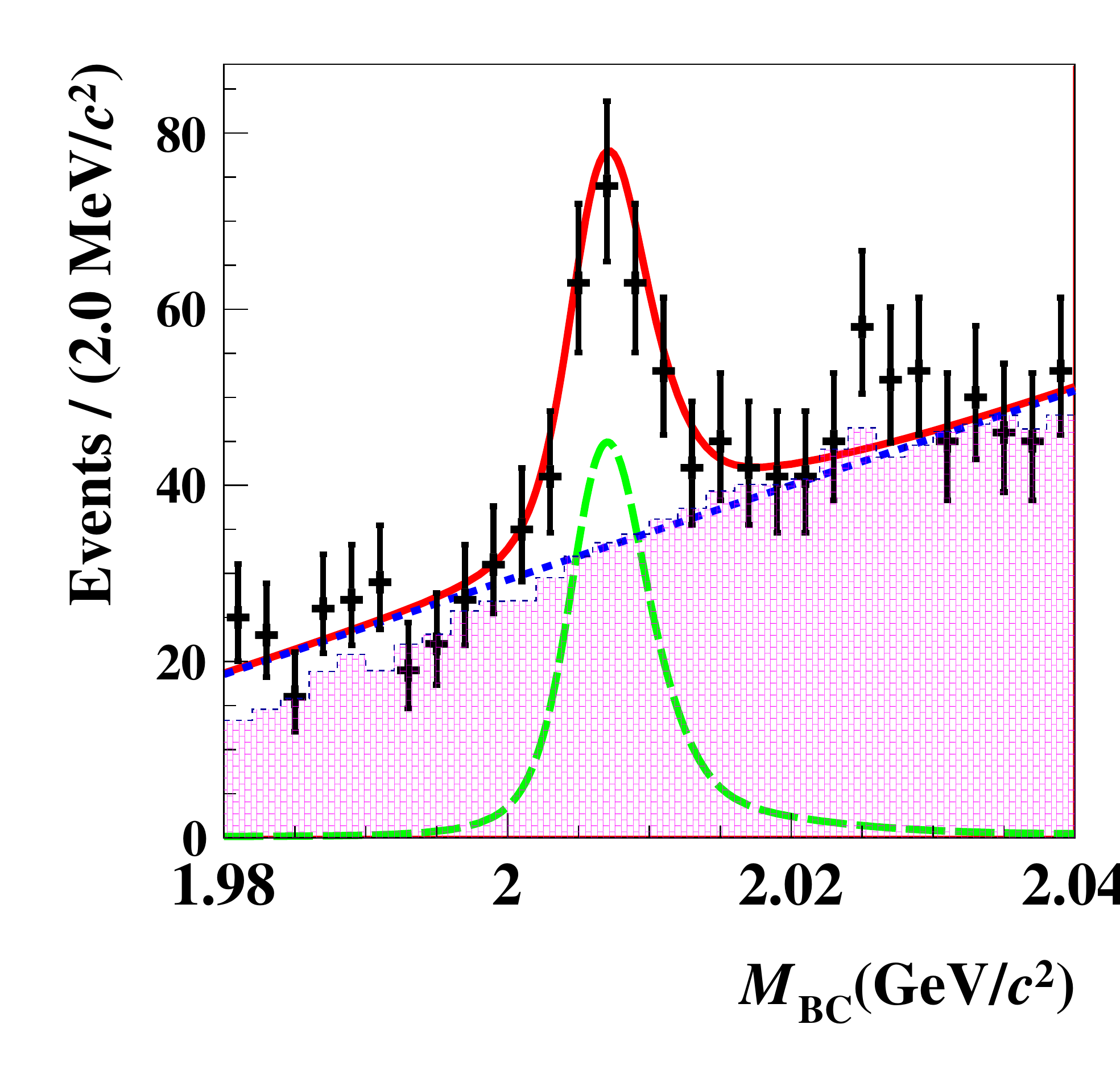}\label{fig:mBC_1}
  \put(-110,100){(b)}
       \includegraphics[width=0.32\textwidth,height=0.2\textheight]{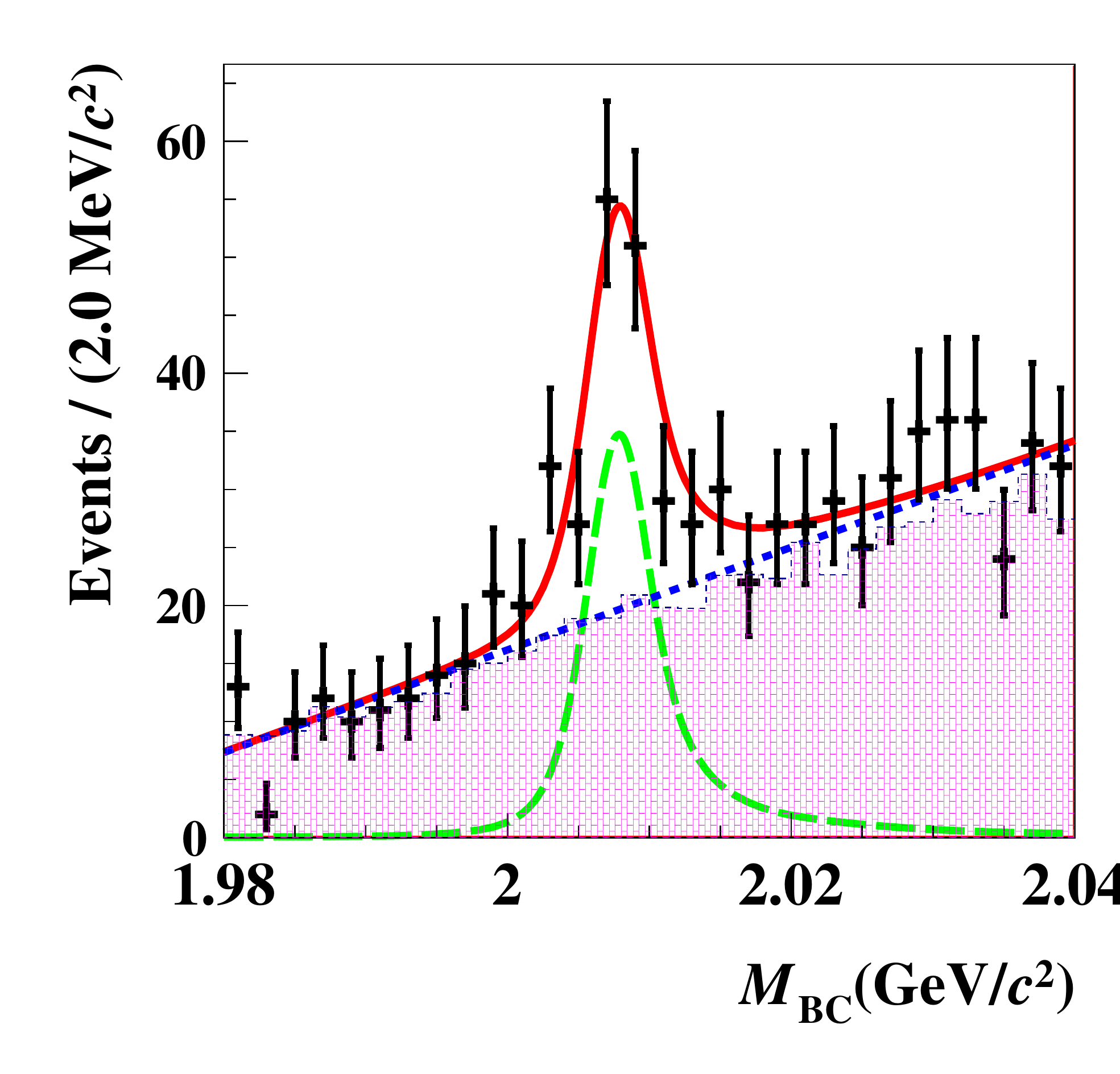}\label{fig:mBC_2}
  \put(-110,100){(c)}
       
        \includegraphics[width=0.32\textwidth,height=0.2\textheight]{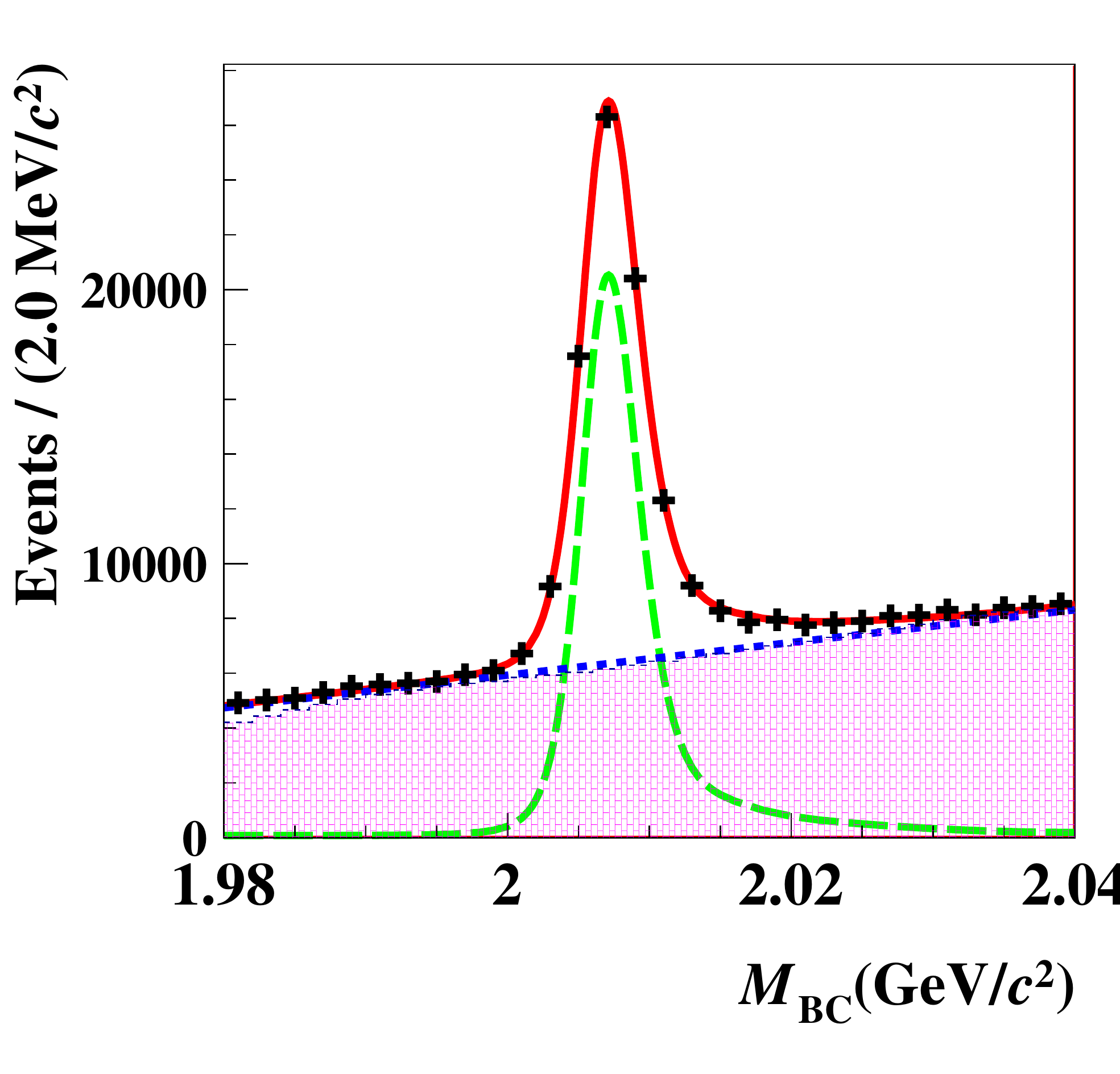}\label{fig:ref_mBC_0}   
     \put(-110,100){(d)}
        \includegraphics[width=0.32\textwidth,height=0.2\textheight]{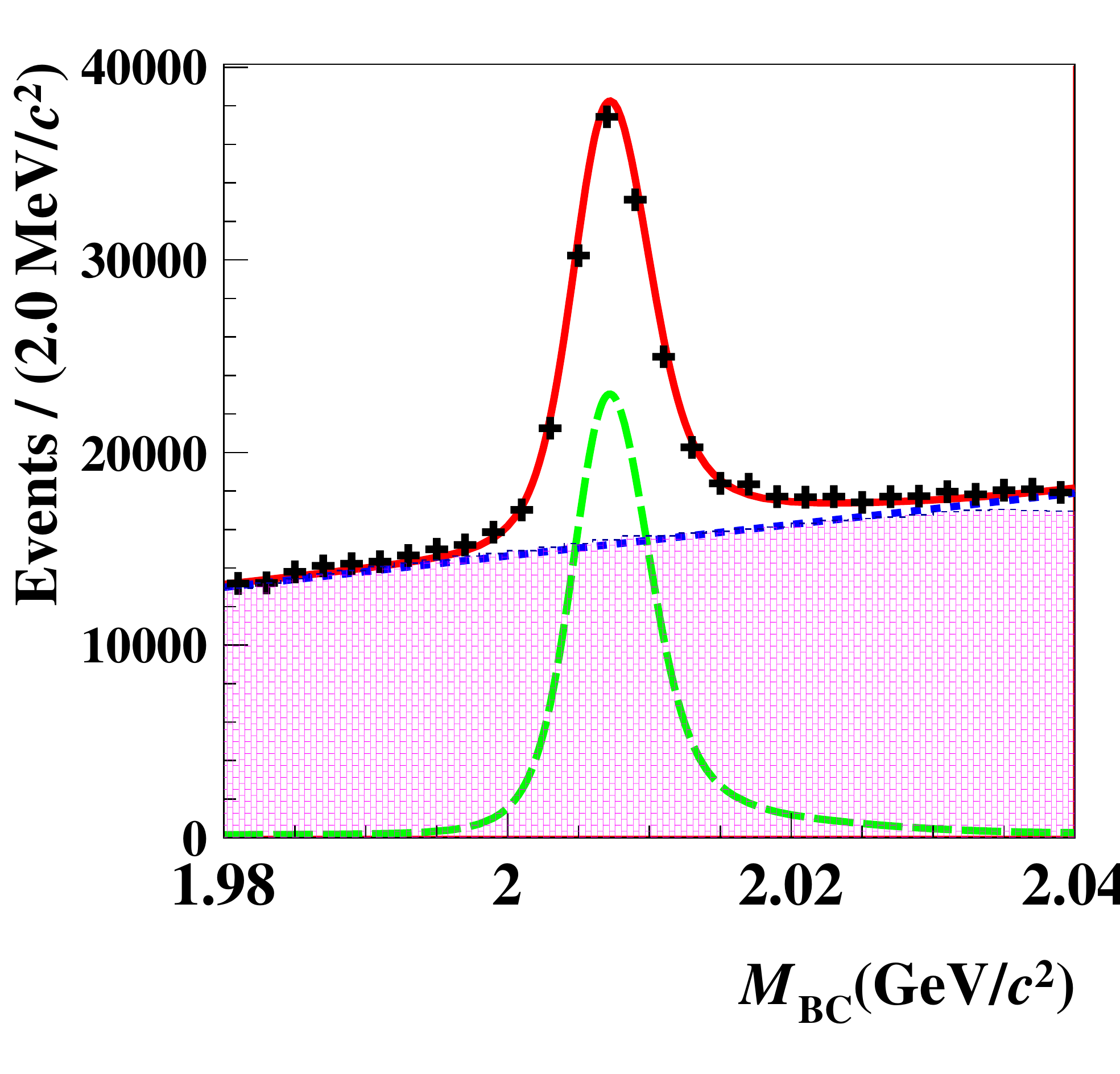}\label{fig:ef_mBC_1}   
  \put(-110,100){(e)}   
        \includegraphics[width=0.32\textwidth,height=0.2\textheight]{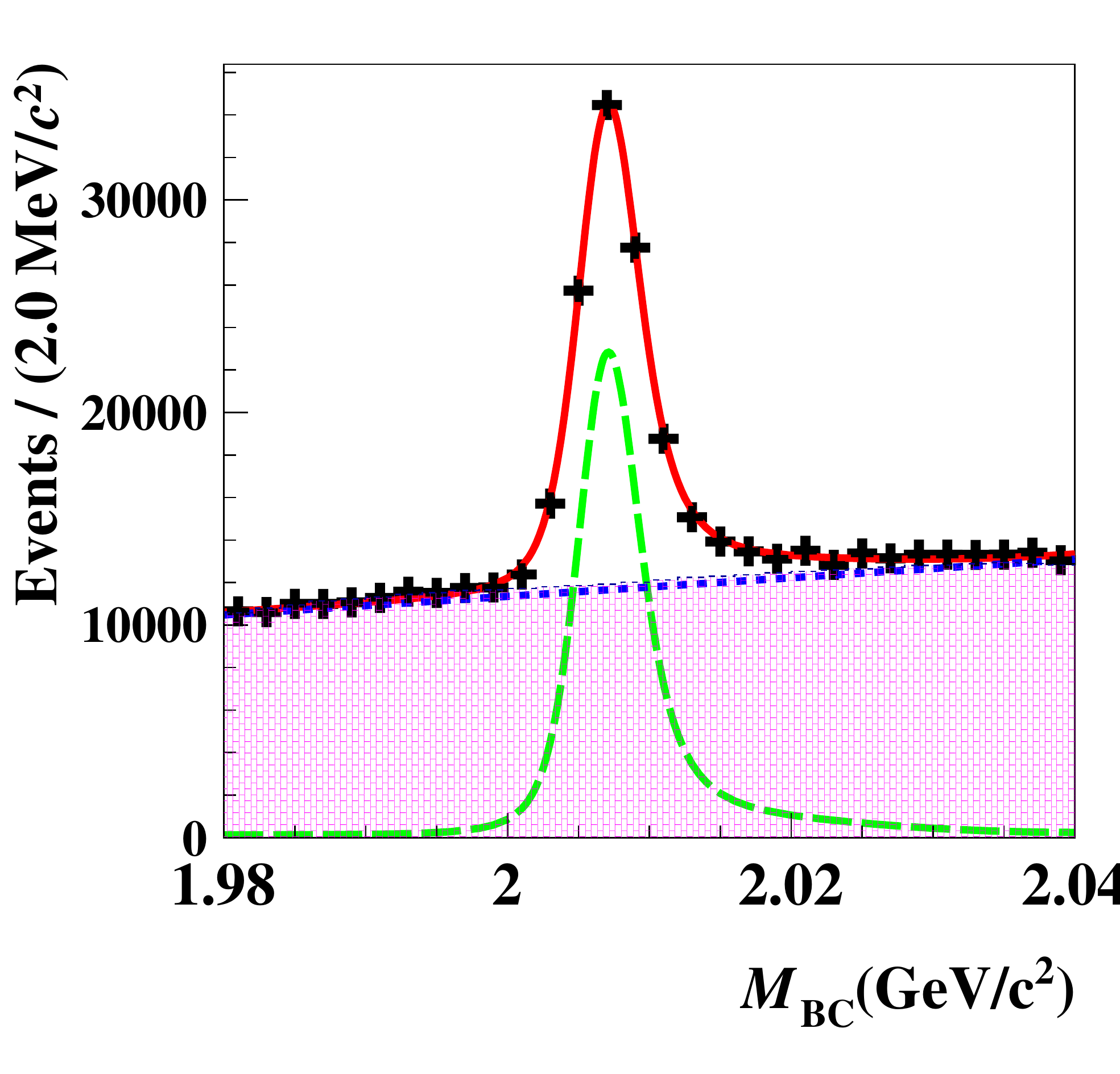}\label{fig:ef_mBC_2}
  \put(-110,100){(f)}
        \end{center}
    \caption{Fits to the $\mBC$ distributions of the candidates for $D^{*0} \rightarrow D^{0} e^{+} e^{-}$~(top) and  $D^{*0} \rightarrow D^{0} \gamma$~(bottom), plots (a) and (d) are reconstructed using $D^{0}\rightarrow K^{-}\pi^{+}$ mode, plots (b) and (e) are reconstructed using $D^{0}\rightarrow K^{-}\pi^{+}\pi^{0}$ mode. Plots (c) and (f) are reconstructed using $D^{0}\rightarrow K^{-}\pi^{+}\pi^{-}\pi^{+}$ mode. Black dots with error bars are data, the solid red lines are total fits, the long-dashed green lines are signals and the dotted blue lines are smooth backgrounds. The shaded areas represent the MC-predicted backgrounds.} 
 \label{fig:fitting_comeb}
\end{figure*}

The individual yields of the signal and reference processes are obtained by simultaneous fits to the \mBC{} distributions. The $M_{\rm BC}$ distributions of the accepted candidates for both signal and reference channels are shown in \figurename~\ref{fig:fitting_comeb}. 
In the fits, the signal shapes are obtained from the corresponding simulated shapes convolving with Gaussian functions to compensate the resolution difference between data and MC simulation. A common $R_{ee}$ is used for all three tag modes in the simultaneous fit, after the relevant efficiencies and branching fractions for each $D^0$ mode are taken into account. 
Note that peaking contributions from the corresponding doubly Cabibbo suppressed modes $\bar{D}^{0} \rightarrow K^{-} \pi^{+}$, $K^{-} \pi^{+}\pi^{0}$ and $K^-\pi^+\pi^-\pi^+$ cancel out in calculating $R_{ee}$. 
Background studies indicate  no peaking structures in the fit range, as illustrated in \figurename~\ref{fig:fitting_comeb}. 
Hence, a second-order polynomial function is used to describe the combinatorial backgrounds. The final fit results are also shown in \figurename~\ref{fig:fitting_comeb},
and the numerical results of the efficiencies and yields are listed in Table~\ref{tab:results}.
The ratio $R_{ee}$  is determined to be $(11.08\pm0.76)\times 10^{-3}$, where the uncertainty is statistical. The statistical significance is estimated to be $13.2\sigma$. The $R_{ee}$ values obtained separately in the three $D^0$ tag modes are consistent. By using the world average value of the branching fraction of \DstarDg{}, the branching fraction of \DstarDee{} is calculated to be $(3.91\pm0.27)\times 10^{-3}$.

\section{Systematic uncertainty}

The main sources of systematic uncertainties are summarized 
in Table~\ref{tab:Error} and discussed in detail next.  
Systematic uncertainties in $D^{0}$ reconstruction are canceled in the ratio $R_{ee}$.

\begin{table}[htp] 
\begin{center}
\caption{Systematic uncertainties in measuring $R_{ee}$.} \label{tab:Error}
   \begin{tabular}{c|c}
	\hline\hline
	Source  & Uncertainty ($\%$)\\
\hline
     $e^{\pm}$ tracking & 2.9\\      
      $e^{\pm}$ PID   &1.4 \\
      Photon detection  & 1.0 \\
      Signal Generator & 1.1\\
   Photon conversion rejection & 1.0  \\
      \dE{} requirement &1.1\\
      Signal shape & 1.1\\
      Background shape & 1.7\\       
\hline
       Total & 4.4\\
	\hline\hline
   \end{tabular}
   \end{center}
 \end{table}

To study the efficiencies of the tracking and PID of low-momentum $e^{\pm}$, a control sample of  $e^{+}e^{-}\to \gamma e^{+}e^{-}$ is selected. 
Due to relatively large  efficiency differences, a reweighting procedure according to the two-dimensional kinematic distribution in transverse momentum and $\cos\theta$   is implemented to reweight the MC-determined efficiency of  \DstarDee{}. The overall correction factor is estimated to be $(108.5\pm2.9)\%$ for $e^+e^-$ pair. For $e^{\pm}$ PID, the same control sample and reweighting procedure are used. The corresponding overall correction factor is estimated to be $(94.1\pm1.4)\%$. After correcting the signal efficiencies to data, the uncertainty on the correction factor is assigned as the systematic uncertainties. 

In the reference mode \DstarDg, the uncertainty in the reconstruction efficiency of photon is assigned to be  
1.0$\%$ based on the studies with a control sample of $J/\psi \rightarrow \rho^{0}\pi^{0}$ events~\cite{Ablikim:2010zn}.

In the nominal analysis, the PDG value of $m_\rho$ is used in the form factor of $D^{*0}\to D^0e^+e^-$ in the signal generator. In order to assign an uncertainty, we replace $m_\rho$ with $m_\omega$ in the form factor in Eq.~(\ref{Eq:2}). The resultant change on the efficiency, 1.1\%, is taken as the systematic uncertainty. 

The signal efficiency loss due to vetoing photon conversions ($R_{xy}<2.0$~cm) is studied with  a  control sample of 
$\ee\to \gamma J/\psi$, $J/\psi \rightarrow \pi^{0}\pip\pim$ with $\pi^{0}$ decaying to $\gamma \ee$ final state.
The veto efficiency is found to be lower than that determined in MC simulation by a factor $(96.4 \pm 1.0)\%$. After correcting the signal efficiencies to data, we assign 1.0\% as the corresponding systematic uncertainty.

The efficiencies of the $\Delta E$ requirements are estimated by correcting the variable $\Delta E$ in the MC sample according to
the observed resolution difference between data and MC simulation. The change of the obtained efficiencies of the corrected signal MC samples from 
the nominal efficiencies is taken as systematic uncertainty.
For the fits to the $\mBC$ distributions, the signal shapes are varied from the MC-derived shape convolved with a single-Gaussian function to one using MC simulation convolved with a double-Gaussian function, while the background shape is varied from a second-order polynomial function to a linear function. The resultant changes on the final result are adopted as systematic uncertainties.

\section{Summary}

Based on 3.19 fb$^{-1}$ of $e^{+}e^{-}$ collision data collected at $\sqrt{s}$ =4.178 GeV with the BESIII detector, 
the EM Dalitz decay \DstarDee{} is observed for the first time. The branching fraction of this decay relative to that of $D^{*0}\to D^0\gamma$ is measured to be  $(11.08\pm0.76\pm0.49)\times 10^{-3}$. This result deviates from the ratio of 0.67\% according to Eq.~\eqref{Eq:2} with  $3.5 \sigma$.~Using the world average $\mathcal{B}(\DstarDg{})=(35.3\pm 0.9)\%$~\cite{Zyla:2020zbs}, we determine $\mathcal{B}(\DstarDee)=(3.91\pm0.27\pm0.17\pm0.10)\times 10^{-3}$, where the first uncertainty is statistical, the second is systematic and the third is from the uncertainty of the input $\mathcal{B}(\DstarDg{})$. The obtained distribution of virtual photon 4-momentum transfer square is compatible with the model prediction in Eq.~\eqref{Eq:2}, as shown in \figurename~\ref{fig:com_data_mc}(d). However, the statistics for now is still low for meaningful measurements of the $q^2$-dependent form factor. Better precisions can be achieved with more data taken at BESIII in the future~\cite{BESIII:2020nme}, which could be used for more physics goals, such as searching for the light vector boson $X17$ decaying into the $\ee$ pairs~\cite{Castro:2021gdf}.

\acknowledgments{We thank Jia-Jun~Wu for useful discussions. The BESIII collaboration thanks the staff of BEPCII
and the IHEP computing center for their strong support. This work is supported in part by National
Key Research and Development Program of China under Contracts Nos. 2020YFA0406400, 2020YFA0406300;
National Natural Science Foundation of China (NSFC)
under Contracts Nos. 11875115, 11625523, 11635010,
11735014, 11822506, 11835012, 11935015, 11935016,
11935018, 11961141012, 12005311; The Fundamental Research Funds for the Central Universities, Sun Yat-sen University; The Chinese Academy of
Sciences (CAS) Large-Scale Scientific Facility Program;
Joint Large-Scale Scientific Facility Funds of the
NSFC and CAS under Contracts Nos. U1732263,
U1832207, U2032110; CAS Key Research Program of
Frontier Sciences under Contracts Nos. QYZDJ-SSW-SLH003, QYZDJ-SSW-SLH040; 100 Talents Program
of CAS; INPAC and Shanghai Key Laboratory for
Particle Physics and Cosmology; ERC under Contract
No. 758462; European Union Horizon 2020 research and innovation programme under Contract No.
Marie Sklodowska-Curie grant agreement No 894790;
German Research Foundation DFG under Contracts Nos.
443159800, Collaborative Research Center CRC 1044,
FOR 2359, FOR 2359, GRK 214; Istituto Nazionale
di Fisica Nucleare, Italy; Ministry of Development of
Turkey under Contract No. DPT2006K-120470; National
Science and Technology fund; Olle Engkvist Foundation
under Contract No. 200-0605; STFC (United Kingdom);
The Knut and Alice Wallenberg Foundation (Sweden)
under Contract No. 2016.0157; The Royal Society,
UK under Contracts Nos. DH140054, DH160214; The
Swedish Research Council; U. S. Department of Energy
under Contracts Nos. DE-FG02-05ER41374, DE-SC-001206.}



\begin{thebibliography}{90}
\bibitem{Landsberg:1986fd}
L.~G.~Landsberg,
Phys. Rep. \textbf{128}, 301 (1985).

\bibitem{Wu:2019adv}
J.~J.~Wu, T.~S.~H.~Lee and B.~S.~Zou,
Phys. Rev. C \textbf{100}, 035206 (2019).

\bibitem{Achasov:2008zz}
M.~N.~Achasov, K.~I.~Beloborodov, A.~V.~Berdyugin, A.~G.~Bogdanchikov, A.~D.~Bukin, D.~A.~Bukin, A.~V.~Vasilev, V.~B.~Golubev, T.~V.~Dimova and V.~P.~Druzhinin, \textit{et al.}
J. Exp. Theor. Phys. \textbf{107}, 61 (2008).

\bibitem{Anastasi:2016qga}
A.~Anastasi \textit{et al.} (KLOE-2 Collaboration),
Phys. Lett. B \textbf{757}, 362 (2016).


\bibitem{Babusci:2014ldz}
D.~Babusci \textit{et al.} (KLOE-2 Collaboration),
Phys. Lett. B \textbf{742}, 1 (2015).


\bibitem{Ablikim:2018xxs}
M.~Ablikim \textit{et al.} (BESIII Collaboration),
Phys. Lett. B \textbf{783}, 452 (2018).


\bibitem{Ablikim:2017kia}
M.~Ablikim \textit{et al.} (BESIII Collaboration),
Phys. Rev. Lett. \textbf{118}, 221802 (2017).

\bibitem{BESIII:2018qzg}
M.~Ablikim \textit{et al.} (BESIII Collaboration),
Phys. Rev. D \textbf{99},  012006 (2019). 




\bibitem{CroninHennessy:2011xp}
D.~Cronin-Hennessy \textit{et al.} (CLEO Collaboration),
Phys. Rev. D \textbf{86}, 072005 (2012).


\bibitem{Ablikim:2009aa}
M.~Ablikim \textit{et al.} (BESIII Collaboration),
Nucl. Instrum. Methods Phys. Res., Sect. A \textbf{614}, 345 (2010).


\bibitem{Yu:IPAC2016-TUYA01}
   C.~H.~Yu {\it et al.},
  Proceedings of IPAC2016, Busan, Korea (JACoW, Geneva, Switzerland, 2016).
  
  


\bibitem{BESIII:2020nme}
M.~Ablikim \textit{et al.} (BESIII Collaboration),
Chin. Phys. C \textbf{44}, 040001 (2020).

\bibitem{etof}
 X.~Li {\it et al.}, Radiat. Detect. Technol. Methods {\bf 1}, 13 (2017);
 Y.~X.~Guo {\it et al.}, Radiat. Detect. Technol. Methods {\bf 1}, 15 (2017);
 P.~Cao {\it et al.}, Nucl.\ Instrum.\ Methods  Phys. Red., Sect. \ A {\bf 953}, 163053 (2020).



\bibitem{geant4}
  S.~Agostinelli {\it et al.} (GEANT4 Collaboration),
  Nucl.\ Instrum.\ Methods  Phys. Red., Sect. \ A {\bf 506}, 250 (2003).

\bibitem{ref:kkmc}
  S.~Jadach, B.~F.~L.~Ward and Z.~Was,
  Phys.\ Rev.\ D {\bf 63}, 113009 (2001);
  Comput.\ Phys.\ Commun.\  {\bf 130}, 260 (2000).

\bibitem{ref:evtgen}
  D.~J.~Lange,
  Nucl.\ Instrum.\ Methods  Phys. Red., Sect. \ A {\bf 462}, 152 (2001);
  R.~G.~Ping,
  Chin. Phys. C {\bf 32}, 599 (2008).

\bibitem{Zyla:2020zbs}
P.~A.~Zyla \textit{et al.} (Particle Data Group),
Prog. Theor. Exp. Phys. \textbf{2020},  083C01 (2020).


\bibitem{ref:lundcharm}
  J.~C.~Chen, G.~S.~Huang, X.~R.~Qi, D.~H.~Zhang and Y.~S.~Zhu,
  Phys.\ Rev.\ D {\bf 62}, 034003 (2000);
  R.~L.~Yang, R.~G.~Ping and H.~Chen,
  Chin.\ Phys.\ Lett.\  {\bf 31}, 061301 (2014).

\bibitem{photos}
  E.~Richter-Was,
  Phys.\ Lett.\ B {\bf 303}, 163 (1993).



\bibitem{Tan:2021clg}
Y.~Tan, Z.~Zhang and X.~Zhou,
arXiv:2111.04932 .


\bibitem{Xu:2012xq}
Z.~R.~Xu and K.~L.~He,
Chin. Phys. C \textbf{36}, 742 (2012).

\bibitem{Ablikim:2010zn}
M.~Ablikim \textit{et al.} (BESIII Collaboration),
Phys. Rev. D \textbf{81}, 052005 (2010).

\bibitem{Castro:2021gdf}
G.~L.~Castro and N.~Quintero,
Phys. Rev. D \textbf{103}, 093002 (2021).




\end{thebibliography}
\end{document}